\documentclass[11pt]{article}
\usepackage{latexsym,amsmath,units,amssymb,pstricks}
\usepackage[square,comma,sort&compress,numbers]{natbib}
\renewcommand{\a}{{\alpha}}
\renewcommand{\b}{{\beta}}
\newcommand{\ga}{{\gamma}}

\newcommand{\dl}{{\delta}}

\newcommand{\m}{{\mu}}
\newcommand{\n}{{\nu}}
\newcommand{\pa}{{\partial}}

\newcommand{\sig}{{\sigma}}
\newcommand{\Sig}{{\Sigma}}
\renewcommand{\r}{{\rho}}
\renewcommand{\th}{{\theta}}

\newcommand{\la}{{\lambda}}

\newcommand{\om}{{\omega}}
\newcommand{\Om}{{\Omega}}
\newcommand{\imu}{{\rm i}}
\newcommand{\ag}{\mathfrak{g}}

\newcommand{\as}{\mathfrak{s}}
\newcommand{\al}{\mathfrak{l}}

\newcommand{\au}{\mathfrak{u}}
\newcommand{\mG}{\textnormal{\sffamily{\textit{G}}}}
\newcommand{\mH}{\textnormal{\sffamily{\textit{H}}}}
\newcommand{\mJ}{\textnormal{\sffamily{\textit{J}}}}
\newcommand{\mB}{\textnormal{\sffamily{\textit{B}}}}
\newcommand{\mA}{\textnormal{\sffamily{\textit{A}}}}

\numberwithin{equation}{section}
\renewcommand{\theequation}{\arabic{section}.\arabic{equation}}

\begin{document}
\thispagestyle{empty}

\begin{center} {\huge\bf Geometric construction of D-branes\\[9pt]
 in WZW models}
\vskip 24pt 

G. Horcajada and F. Ruiz Ruiz
\vskip 3pt
\emph{Departamento de F\'{\i}sica Te\'orica I, Universidad Complutense de 
       Madrid \\ 28040 Madrid, Spain}

\vskip 36pt 
\today

\vskip 18pt
\begin{abstract}
  The geometric description of D-branes in WZW models is pushed forward. Our
  starting point is a gluing condition\, $\mJ_{+\!}=F\mJ_-$ that matches the
  model's chiral currents at the worldsheet boundary through a linear map $F$
  acting on the WZW Lie algebra. The equivalence of boundary and gluing
  conditions of this type is studied in detail. The analysis involves a
  thorough discussion of Frobenius integrability, shows that $F$ must be an
  isometry, and applies to both metrically degenerate and nondegenerate
  D-branes. The isometry $F$ need not be a Lie algebra automorphism nor
  constantly defined over the brane. This approach, when applied to isometries
  of the form $F=R$ with $R$ a constant Lie algebra automorphism, validates
  metrically degenerate $R$-twined conjugacy classes as D-branes.  It also
  shows that no D-branes exist in semisimple WZW models for constant\, $F=-R$.
\end{abstract}
\end{center}

\vspace{60pt} {\sc Keywords:} D-branes, Bosonic Strings

\newpage
\vskip 36pt
\section{Introduction}

D-branes have become one of the main research topics in the string literature
since the mid nineties. There are many reasons for this. Among them, the
evidence that D-branes provide soliton and bound states in string
backgrounds~\cite{Witten-bound} and the realization that they become upon
quantization noncommutative
spacetimes~\cite{Chu-Ho-flat,Schomerus,Seiberg-Witten, Chu-Ho-pp, ARS-fuzzy,
  Horcajada-Ruiz}.

Since WZW models are the building blocks of many string
backgrounds~\cite{Rahmfeld,Berkovits-1,Berkovits-2}, one sensible program to
study D-branes and their properties is to consider their occurrence in models
of this type. In fact, there are various approaches to D-branes in WZW models.
Among them, the geometric approach~\cite{Alekseev-Schomerus, Stanciu-3,
  Stanciu-manifolds, Stanciu-note, FS-more, Bachas-Petropoulos,
  Ribault-Schomerus, Hikida, Fredenhagen-Quella,Cheung-Freidel, HHRR}, that
regards D-branes as spacetime's submanifolds on which the string worldsheet
boundary may be embedded, and the algebraic approach~\cite{Birke, Fuchs,
  Felder, Schomerus-Saleur, Gotz, Saleur, Quella, Creutzig}, that makes use of
boundary conformal field theory.

In this paper we reexamine the geometric description of D-branes in a WZW
model. The definition of D-brane that we will be using is the na\"ive
geometric one; see Section 2 for details. A D\emph{p}-brane in a string
background\, $(\mG_{\m\n},\mH_{\m\n\r})$ \,is any $(p+1)$-dimensional
submanifold~$N$ containing all possible motions for the string
endpoints. These motions are specified by the boundary conditions for the
string, which in turn can be viewed as a system of first order differential
equations characterized by a two-form $\om$ globally defined\footnote{We will
  use $\mB$ for two-forms locally defined on the whole group manifold such
  that $d\mB=\mH$, and the Greek letter~$\om$ for the two-form globally
  defined on the submanifold $N$.} on $N$ such that\,
\hbox{$d\om=\mH\big\vert_N$}.  A way to construct D-branes is thus to find all
two-forms $\om$ for which the boundary conditions can be integrated.

Our starting point for the geometric characterization of D-branes in a WZW
model is a condition\, $\mJ_+=F\mJ_-$, called gluing condition, that matches
the model's chiral currents $\mJ_-$ and $\mJ_+$ at the world sheet boundary
through a linear map $F$ that acts on the model's Lie algebra.  This matching
condition is not a boundary condition, for it is not obtained by setting to
zero the boundary term that arises from the variation of the model's classical
action. However, it does specify, for every linear map $F$, vector fields
characterizing tangent motions of the string endpoints. If these vector fields
define an integrable distribution, they span the tangent bundle of a
submanifold~$N$ of the spacetime group manifold.  The submanifold $N$ is a
D-brane if the gluing condition can be written as a boundary condition with a
two-form $\om$ globally defined on $N$ such that\, $d\om=\mH\big\vert_N$.

We cross examine this approach for WZW models with arbitrary real Lie group
$G$. Our only assumption is that the corresponding Lie algebra admits an
invariant nondegenerate metric $\Om$.  This includes in particular noncompact
group manifolds with Lorentzian signature, for which there exist metrically
degenerate submanifolds~$N$ such that the tangent space $T_gG$ at any point
$g$ in $N$ cannot be written as an orthogonal sum $T_gN\oplus\/T_gN^\bot$.

If $F$ is a constant $\Om$-preserving Lie algebra automorphism and the
orthogonal decomposition\, $T_gG=T_gN\oplus\/T_gN^\bot$\, is assumed, the
vector fields defined by $F$ are known to be integrable and the two-form $\om$
satisfying\, $d\om=\mH\big\vert_N$ \,is well known~\cite{Alekseev-Schomerus,
  Stanciu-3, Stanciu-manifolds, Stanciu-note}.  Our interest is in cases
escaping these two assumptions. In this more general setting, the situation is
very different. Firstly, because for an arbitrary linear map $F(g)$, the
vector fields specified by the gluing condition, call them $t_i$, do not
always define an integrable distribution. And secondly, because even if they
do, the corresponding gluing condition cannot always be written as a boundary
condition with a two-form $\om$ globally defined on $N$ such that
$d\om=\mH\big\vert_N$.  In this regard, we prove the following two
  results. Every boundary condition for a D-brane $N$ can be written as a
  gluing condition\, $\mJ_+=F\mJ_-$, provided\,
  \hbox{$\textnormal{det}\,(\mG\big\vert_{N\!}-\om)\neq\/0$}, where
  $\mG\big\vert_N$ is the induced metric on $N$. And every gluing condition can
be written as a boundary condition if the linear map $F(g)$ is an isometry of
$\Om$, in which case the two-form $\om$ exists globally and is uniquely
defined by its action $\om(t_i,t_j)$ on the vector fields $t_i$ defined by
$F(g)$.  For a general isometry $F(g)$, the requirement\,
$d\om=\mH\big\vert_N$ \,however does not hold but it becomes a matter of
straightforward algebra to check it in every instance.
These two results open some problems, among them studying the
  applicability of this approach to D-branes for which a full set of gluing
  conditions is not known~\cite{Fredenhagen-Quella,Quella-2}.

The paper is organized as follows. Section 2 poses the problem and reviews the
description of D-branes in WZW models in terms of the gluing condition for the
chiral currents. In Section~3, integrability in terms of Frobenius theorem is
studied and it is shown that the two-form $\om$ for which the gluing condition
becomes a boundary condition exists if and only if $F(g)$ is an isometry of
$\Om$. Section 4 contains a discussion of the limitations of the gluing
condition approach. Isometries of the form $F=\pm\/R$, where $R$ is a constant
Lie algebra automorphism, are considered in Section5. The case $F=R$ has been
studied by other Authors~\cite{Alekseev-Schomerus, Stanciu-3,
  Stanciu-manifolds, Stanciu-note} under the hypothesis that\,
$T_gG=T_gN\oplus\/T_gN^\bot$, the resulting \hbox{D-branes} being $R$-twined
conjugacy classes. It is shown that this result holds even if the latter
assumption on $T_gG$ fails.  As regards the case $F=\!-R$, it is proved that,
contrarily to some claims, the gluing condition for $F=\!-R$ does not provide
D-branes for semisimple Lie algebras. In Section~6, some examples of
$g$-dependent isometries $F(g)$ are considered. It is shown that two different
isometries may define the same integrable distribution but different two-forms
$\om$, one of them satisfying $d\om=\mH\big\vert_N$, hence defining a D-brane,
and the other one not. We close the paper with our conclusions and three short
appendices collecting technical points.

\section{Gluing conditions for chiral currents}

In the sigma model approach, a D\emph{p}-brane in a string background
$(\mG_{\!\m\n},\mH_{\!\m\n\la})$ is a $(p\!+1)$-dimensional submanifold $N$ on
which the endpoints of an open string may lie. The submanifold $N$ has
embedded coordinates $\,x^\m(\tau)=X^\m(\tau,\sig)\big\vert_{\pa\Sig}$ and
these must satisfy the boundary conditions
\begin{equation}
  \big(\pa_i f^\m \mG_{\!\m\n}\,\pa_\sig X^\n 
  - \om_{\!ij}\,\pa_\tau \a^j\big)\, \Big\vert_{\pa\Sig} =0
  \qquad i=1,\ldots,p+1 \,.
  \label{BC-sigma}
\end{equation}
Here $\,\a^1,\ldots,\a^{p+1}$ are local coordinates on the D\emph{p}-brane, so
that $\,x^\m\! = f^\m(\a^1,\ldots,\a^{p+1})$, and $\om_{ij}$ are the
components of a two-form $\om$ globally defined on the brane such that\,
$d\om=\mH\big\vert_N$.

We are interested in \hbox{D-branes} in string backgrounds described by WZW
models~\cite{Witten} with real Lie group $G$ and Lie algebra $\ag$, both of
dimension ${\tt d}$. The Lie algebra $\ag$ is a vector space over ${\bf R}$
and has generators $\{T_A\}$ with commutation relations
\begin{equation}
   [\,T_A,T_B\,] = f_{AB}{\!}^C\, T_C\qquad A,B,C=1,\ldots, {\tt d}\,.
\label{CR}
\end{equation}
The algebra $\ag$ is assumed to have a nondegenerate invariant metric
$\Om$, of arbitrary signature, with
components $\,\Om_{AB}=\Om(T_A,T_B)$, so that 
\begin{equation}
 \Om\big(\,[T_A,T_B],\,T_C\big) = \Om\big(T_A,\,[T_B,T_C]\,\big) 
 ~~\Leftrightarrow~~ 
   f_{AB}{\!}^D\,\Om_{DC} = \Om_{AD}\,f_{BC}{\!}^D \,.
\label{bilinear}
\end{equation}
The existence of such a metric is the only restriction on $\ag$. The group $G$
is taken as the connected component obtained from $\ag$ through
exponentiation.

If $X^\m$ are local coordinates in $G$, the left-invariant~\,$e^A{\!}_\m$ and
right-invariant~\,$\bar{e}^{A}{\!}_\m$ vielbeins that map the group $G$ to its
tangent space $T_gG$ at $g$ are defined by
\begin{equation*}
    g^{-1}\,dg = T_A \,e^A{\!}_\m\,dX^\m \qquad 
    dg\,g^{-1} = T_A \,\bar{e}^A{\!}_\m\,dX^\m\,.
\end{equation*}
In terms of them, the adjoint action of the group on the Lie algebra is
\begin{equation}
   {\rm Ad}_g (T_A) =  g\,T_A\,g^{-1\!} 
         = T_B\,\bar{e}^{\,B}{\!}_\m\> (e^{-1})^\m{\!}_A
             ~~~~\Leftrightarrow~~~~  {\rm Ad}_g= \bar{e}\,e^{-1}\,.
\label{Adg}
\end{equation}
The string background $(\mG_{\!\m\n},\mH_{\!\m\n\la})$ is defined from $\Om$ by
\begin{eqnarray}
   & \mG_{\!\m\n} = \Om\,\big(g^{-1}\pa_\m g\,,\,g^{-1}\pa_\n g\big) & 
    \nonumber \\[3pt]
   & \mH_{\!\m\n\la} = \Om\, \big(\big[g^{-1} \pa_\m g\,,
                \,g^{-1} \pa_\n g\big]\,,\,g^{-1} \pa_\la g\big)\,. &
\label{H}
\end{eqnarray}
By construction, the metric $\mG_{\!\m\n}$ is bi-invariant, 
\begin{equation}
     \mG_{\!\m\n}= e^A{\!}_\m\,\Om_{AB}\,e^B{\!}_\n 
            = \bar{e}^{\,A}{\!}_\m\,\Om_{AB}\,\bar{e}^{\,B}{\!}_\n
   ~~\Leftrightarrow~~ 
   \mG=e^{\rm T}\Om\,e= \bar{e}^{\rm T}\Om\,\bar{e}\,,
\label{bi-invariance}
\end{equation}
the superscript $\textnormal{T}$ denoting transposition.  In this paper the
standard notation\, $\mG(a,b)=\mG_{\!\m\n}a^\m b^\n$ \,will be used.

The WZW classical action for the open string in the background
$(\mG_{\!\m\n},\mH_{\!\m\n\la})$ can be written as~\cite{Klimcik-Severa}
\begin{equation}
  S_{\textnormal{\sc wzw}} = \frac{k}{4\pi} \int_\Sig d\hspace{.7pt}^{2\!}\sig~
      \Om\,\big(g^{-1}\pa_ ag,g^{-1}\pa^a g\big) 
    + \frac{k}{4\pi}\>\bigg( \int_\Sig g^*\mB
            + \int_{\pa\Sig} g^*\mA \bigg)\,,
\label{classical}
\end{equation}
with $g=g(X^\m(\tau,\sig))$ and $\sig^a=(\tau,\sig)$ world sheet indices.
Here $\mB$ is any two-form defined on $G$ such that $\,\mH=d\mB$, and $g^*\mB$
is its pullback.  The form $\mB$ may not be globally defined, but must exist
locally. This is the case, for example, if $\mH$ is not exact. The one-form
$\mA$ is defined on the D-brane, exists at least locally and is such that
$d\mA=\om-\mB\big\vert_N$. See ref.~\cite{Klimcik-Severa} for details.

In worldsheet coordinates \hbox{$\,\sig^\pm \!= \tau\pm \sig$}, the
  field equations read\, $\pa_+\mJ_-\! =\pa_- \mJ_+\!=0$, where the chiral
  currents $\mJ_-$ and $\mJ_+$ are given by
\begin{equation*}
        \mJ_-(\sig^-)=g^{-1}\pa_-g\qquad \mJ_+(\sig^+)=-\,\pa_+g\,g^{-1}\,.
\end{equation*}
Due to the simplicity of the solutions for $\mJ_+$ and $\mJ_-$, we are
interested in formulating the boundary conditions for a D-brane in terms of
$\mJ_+$ and $\mJ_-$. We will then assume that there exists a mapping $F$ from
$\ag$ to $\ag$ relating the two currents at the world sheet boundary, that
is, $\mJ_+\! = F(\mJ_-)$ \,at\, $\sig^+\!= \sig^-$. Recalling that
\hbox{D-branes} in the sigma model approach are defined by boundary conditions
of order one in $\pa_\pm X^\m\big\vert_{\pa\Sig}$ and noting that $\mJ_-$ and
$\mJ_+$ are already order one in $\pa_\pm X^\m$, we restrict ourselves to
linear maps $F(g)$ that may depend on $g$ but not on $\mJ_\pm(g)$. For the
chiral currents at a D-brane we thus require
\begin{equation}
  \mJ_+\Big\vert_{\sig^+\!= \sig^-} = F(g) \mJ_- \Big\vert_{\sig^+\!= \sig^-}\,,
\label{GC-general}
\end{equation}
with $F(g)$, for every $g$ in $N$, a linear map that acts on $\ag$ as a vector
space. The linear map $F(g)$ is represented by a real ${\tt d}{\sc \times}{\tt
  d}$ matrix with entries $F^A{\!}_B$ given by $F(g)T_B=T_A\,F^A{\!}_B(g)$.
It is important to note that eq.~(\ref{GC-general}) is not a boundary
condition derived from the classical action above but a working hypothesis
formulated ad hoc.  To keep this in mind, eq.~(\ref{GC-general}) is called
gluing condition, rather than boundary condition. We will take it as starting
point for the construction of D-branes.

Eq.~(\ref{GC-general}) defines a D-brane if it can be written as a sigma model
boundary condition~(\ref{BC-sigma}), with $\om$ a two-form globally defined on
$N$ such that~\cite{Klimcik-Severa}
\begin{equation}
    \mH\big\vert_N  = d\om  \,.
\label{H-condition}
\end{equation}
We will show in Section 3 that every boundary condition~(\ref{BC-sigma}) can
be written as a gluing condition~(\ref{GC-general}), with $F$ and isometry of
$\Om$, except for D-branes $N$ such that\,
$\textnormal{det}\,(\mG\big\vert_{N\!}-\om)=0$, where $G\big\vert_N$ is the
induced metric on $N$. Apart from these instances, the gluing condition is
capable of constructing all D-branes defined by boundary conditions.

It is convenient to write the gluing condition~(\ref{GC-general}) in terms of
local coordinates~$X^\m$. To do this~\cite{Stanciu-manifolds}, it is enough to
use for the chiral currents their expressions
\begin{equation*}
    J_- = \pa_-X^\m\>e^A{\!}_\m\,T_A \qquad 
    J_+ = -\,\pa_+X^\m\>\bar{e}^A{\!}_\m\,T_A
\end{equation*}
in terms of the string coordinates $X^\m$ and the left and right-invariant
vielbeins $\,e^{\,A}{}_\m$ and $\bar{e}^{\,A}{}_\m$.  This yields 
\begin{equation}
  \pa_+ X^\m \Big\vert_{\pa\Sig} 
    = {\cal F}^\m{\!}_\n\, \pa_- X^\n \, \Big\vert_{\pa\Sig} \,,
\label{GC-pm}
\end{equation}
where the matrix ${\cal F}$ is defined by
\begin{equation}
   {\cal F}(x)= -\, \bar{e}^{\,-1} F(g)\,e \,.
\label{calF}
\end{equation}
In worldsheet coordinates $\tau$ and $\sig$, eq.~(\ref{GC-pm}) takes the form
\begin{equation}
  \big( {\cal F} - 1 \big)\,\pa_\tau X \big\vert_{\pa\Sig}
  = \big( {\cal F} + 1 \big)\, \pa_\sig X \big\vert_{\pa\Sig} \,.
\label{GC-ts}
\end{equation}
We emphasize that the matrix~${\cal F}(x)$ is defined on $N$ and depends on
the string endpoints coordinates $x^{\m\!}=X^\m\big\vert_{\pa\Sig\,}$ through
$e(x)$, $\bar{e}(x)$ and $F\big(g(x)\big)$. It however acts on arbitrary
tangent vectors in $T_gG$. To ease the notation, whenever there is no
confusion we will remove from $F(g)$ and ${\cal F}(g)$ the dependence on $g$.

\section{Integration of the gluing condition and D-branes}

In this section, the gluing condition~(\ref{GC-ts}) is explicitly solved for\,
$\pa_\tau\/X^\m\big\vert_{\pa\Sig}$. The solution happens to be given in terms
of vector fields defined by the linear map $F$. The involutivity requisite
that such fields must satisfy to define a foliation of $G$ in terms of a
family of submanifolds $N$ is studied in detail. Finally, it is shown that
the gluing condition for a linear map $F$ takes the form of a boundary
condition if and only if $F$ is an isometry of $\Om$, and the two-form $\om$
is constructed from~$F$. When the resulting $\om$ satisfies
$d\om=\mH\big\vert_N$, the submanifold $N$ is a D-brane.

\subsection{Conditions for the existence of a D-brane}

The set of possible motions of the string endpoints at an arbitrary
spacetime point $g$ is the set $\Pi_g$ of solutions\,
$t^\m(x)\!:=\pa_\tau\/X^\m\big\vert_{\pa\Sig\,}$ \,to the gluing
condition~(\ref{GC-ts}) for some\, $u^\m(x)\!:=\pa_\sig
X^\m\big\vert_{\pa\Sig\,}$. Since\, $t(x)=t^\m(x)\,\pa_\m$ \,and\,
$u(x)=u^\m(x)\,\pa_\m$ \,are tangent vectors to $G$ at $g$, we may write
\begin{equation*}
  \Pi_g=\Big\{\,t\in T_g(G)\!:~ 
       \big[{\cal F}(g)- 1\big] t = \big[{\cal F}(g) + 1\big]u
         ~~\textnormal{for}\,~  u\!\in\! T_g(G)\,\Big\}.
\end{equation*}
Equivalently,
\begin{equation}
 \Pi_g=\big\{\,t\in T_g(G)\!:\, ({\cal F}- 1)\,t
        \in \textnormal{Im}\,({\cal F}+ 1) \big\}\,.
\end{equation}
The set $\Pi_g$ is a linear subspace of the tangent space $T_gG$ and we will
often call it the tangent plane at $g$ defined by $F$.

Consider\, $v\!\in\!\textnormal{Im}\,({\cal F}+1)$, so that there exists $w$
in $T_g(G)$ such that\, $v=({\cal F}+1)\,w$. It follows that\; $({\cal
  F}-1)\,v=({\cal F}+1)\,({\cal F}-1)\,w$ \;belongs to\; $\textnormal{Im}\,({\cal
  F} + 1)$. Hence $v$ is in $\Pi_g$ and 
\begin{equation*}
  \textnormal{Im}\,( {\cal F} + 1)\subset\Pi_g\,.
\end{equation*}
Consider now $v'$ in $\Pi_g$. It then exists $w'$ in $T_gG$ such that\, $({\cal
  F}-1)v'=({\cal F}+1)w'$. This implies that $v'=\frac{1}{2}({\cal F}+1)(v'-w')$,
so that $v'$ belongs to\, $\textnormal{Im}\,({\cal F} + 1)$ \,and
\begin{equation*}
  \Pi_g\subset\textnormal{Im}\,({\cal F} + 1)\,.
\end{equation*}
Hence
\begin{equation}
    \Pi_g= \textnormal{Im}\,( {\cal F} + 1)  \,.
\label{Pi-g}
\end{equation}
An alternative derivation of this result in terms of the eigenvectors of
${\cal F}$ can be found in Appendix~A.

Since the gluing condition~(\ref{GC-ts}) holds for arbitrary $g$, the solution
$t(x)$ defines a vector field for a given $u(x)$. If $M$ is a submanifold of
$G$, we define
\begin{equation}
   \Pi^{M}=\left\{\left(g,\Pi_g\right):~ g\in\/M\,\right\} .
\label{Pi-M}
\end{equation}
$\Pi^{M}$ is a distribution on $M$ if the tangent plane $\Pi_g$ has the same
dimension for all $g$ in $M$. According to Frobenius theorem~\cite{Boothby}, a
distribution $\Pi^M$ is integrable if and only if it is involutive.
Integrability ensures that $\Pi_g$ is, for all $g$ in $M$, not just a tangent
plane but the tangent space to a submanifold $N$ of $M$, that is
$\Pi_g=T_gN$. Involutivity states that the commutator of any two vector
fields $t_1$ and $t_2$ taking values in $\Pi^M$ also takes values in $\Pi^M$,
\begin{equation}
 \left[t_1,t_2\right] (g)\in \Pi_g \,.
\label{Fro}
\end{equation} 
For the manifold $N$ to define a D-brane, it must contain all the points $g$
in $G$ connected by the integral curves of the vector fields $t$. This
condition cannot be relaxed, since one would then leave out from the D-brane
points at which the open string may end. See Section 4 for examples.

As a practical matter, to determine if a linear map $F$ defines a D-brane,
one may proceed in three steps:

{\bf\emph{Step 1.}} Study the rank of the matrix ${\cal F}(g)+1$ as a function
of $g$. Consider a submanifold $D_n(F)$ formed by the points $g$ in $G$ such
that (i) the rank of\, ${\cal F}(g)+1$ \,is $n$, and (ii) $g$ is not connected
by integral curves of the vector fields~$t$ with points~$g'$ at which the rank
of\, ${\cal F}(g')+1$ \,is different from~$n$.

{\bf\emph{Step 2.}} Check the involutivity condition~(\ref{Fro}) in
$D_n(F)$. If it holds, the distribution $\Pi^{D_n(F)}$ is the tangent bundle
of a submanifold $N$ of $G$ of dimension $n$, or more precisely of a family of
submanifolds which foliate $D_n(F)$.

{\bf\emph{Step 3.}} Find a two-form $\om$ globally defined on $N$ for
which the gluing condition for $F$ can be recast as a sigma model boundary
condition and such that $d\om=H\big\vert_N$. If such a $\om$ exists, the
submanifold $N$ is a D-brane of dimension $n$.

  In what follows we further elaborate these three steps.

\subsection{Involutivity in detail}

The definition of ${\cal F}$ in~(\ref{calF}) and the expression for the group
adjoint action in~(\ref{Adg}) imply that\, \hbox{${\cal
    F}+1=e^{\,-1}(-\textnormal{Ad}_{g^{-1}}F+1)\,e$}. The space of tangent
directions $\Pi_g=\textnormal{Im}({\cal F}+1)$ can then be written as
\begin{equation}
   \Pi_g=g\,\big[\,\textnormal{Ad}_{g^{-1}}F(g)-1\,\big]\,\ag\,.
\label{Tg}
\end{equation}
For every $V$ in the Lie algebra $\ag$,
\begin{equation}
   g\,\big[\textnormal{Ad}_{g^{-1}}F(g)-1\big]V=F(g)\,Vg - g\,V
\label{vec-field}
\end{equation}
is a vector field. It is actually the sum of a right-invariant vector field
\hbox{$\,Y_{}g$}, with \hbox{$\,\!Y=F(g)V$}, and a left-invariant vector field
\hbox{$\,gY$}, with \hbox{$\,Y\!=\!-V$}.

Right and left-invariant vector fields act on differentiable functions $f$
defined on $G$ and taking values in ${\bf R}$ according to
\begin{equation}
  Y\!g\, \big(f(g)\big)  
       = \frac{d}{dt}~f\big( e^{tY} g\,\big)\bigg\vert_{t=0}   \qquad
  gY\, \big(f(g)\big) 
        = \frac{d}{dt}~f\big( g\,e^{tY}\, \big)\bigg\vert_{t=0}\,.
\label{left-right}
\end{equation}
If $g$ is parameterized by coordinates $x^\m$, the vector field components of
$\,Y\!g\,$ and $\,gY\,$ are 
\begin{align*}
   Y\!g & = Y^A\,T_A\, g = Y^A\,{(\bar{e}^{\,-1})}^\m{}_A\,\pa_\m
          ={(Yg)}^\m\,\pa_\m  \\[3pt]
   gY\! & = Y^A\,gT_A = Y^A\,{(e^{-1})}^\m{}_A\,\pa_\m={(gY)}^\m\,\pa_\m \,.
\end{align*}
These equations and $V^{A\!}=e^{A}{\!}_\m \,v^\m$ \;provide\; $F(g)Vg -
gV=\!-\,{\big[({\cal F}+1)\,v \big]}^\m\pa_\m$, which again gives for the
vector field $t$ the form used in eq.~(\ref{Pi-g}). Since $\{T_A\}$ is a basis of
$\ag$, the vector fields that define $\Pi_g$ read 
\begin{equation}
      t_A = FT_Ag-gT_A = \big[\,{(\bar{e\,}^{-1})^\m}_B F^B{}_A  
                   - \,{(e^{-1})^\m}_A\,\big]\,\pa_\m \,.
\label{t_A}
\end{equation}
These fields completely determine the motions of the string endpoints
solving the gluing condition with linear map $F$.

The rank of\, ${\cal F}+1$ \,is obviously equal to the rank of\,
$\textnormal{Ad}_{g^{-1}}F(g) -1$. Say it takes the value $n$ for all $g$ in a
domain $D_n(F)$ in $G$. Assume further that the integral curves of the fields
$t_A$ are in $D_n(F)$.  The involutivity condition~(\ref{Fro}) requires that,
for all $U$ and $V$ in $\ag$ and for all $g$ in $D_n(F)$, there exist $W$ in
$\ag$ such that
\begin{equation}
  \big[\,F(g)\,Ug - g\,U\, ,\, F(g)\,V\,g - gV\,\big] 
     =  F(g)\,Wg - gW\,.
\label{involutive}
\end{equation}
It is important to keep in mind that $W$ need not be the same for all
$g$. Eq.~(\ref{involutive}) is an equation in $F(g)$, in the sense that $W$
does not exist for every linear map $F(g)$. After expanding its left hand side,
it becomes
\begin{equation}
  \big[\,F(g)Ug\, , F(g)Vg\big] 
  - \big[\,F(g)Ug\, ,  gV\,\big] - \big[\,gU\, , F(g)Vg\, \big] 
  + \big[\,gU\, , gV\,\big]  =  F(g)\,Wg - gW\,.
\label{invol}
\end{equation}
Let us understand each one of the terms in this expression.  Using
eqs.~(\ref{left-right}), the action of first term on an arbitrary function $f$
is
\begin{equation*}
    \big[ F(g)Ug\,,\,F(g)Vg \big] \>\big(f(g)\big) = 
      \frac{\pa^2}{\pa s\,\pa t}~
           f\Big( e^{\,\displaystyle{ tF( e^{s F(g)U}g)\,V}}\,
                e^{\displaystyle{s F(g) U}}  g \Big)~\Big\vert_{s=t=0}\!
          -\,\big(U\leftrightarrow V\big) \,.
\end{equation*}
After performing the derivatives with respect to $s$ and $t$ and using
eqs.~(\ref{left-right}), this reduces to
\begin{align}
 \big[ F(g)Ug\,, & \,F(g)Vg \big] \>\big(f(g)\big)  = 
     \big[ F(g)V\,,\,F(g)U\big]g\>\big(f(g)\big) \nonumber \\[3pt]
    & +\Big( F(g)Ug \>\big(F(g)\big) \Big)
    ~ Vg \,\big(f(g)\big)
    - \Big( F(g)Vg \>\big(F(g)\big)\Big)
    ~ Ug \,\big(f(g)\big) \,. \label{first-g}
\end{align}
Proceeding similarly for the other commutators in eq.~(\ref{invol}), we obtain
\begin{equation}
    \big[ F(g)Ug\,,gV\big]\>\big(f(g)\big) =
      -\,\Big( gV\>\big(F(g)\big)\Big)~ Ug\> \big(f(g)\big)
\label{second-g}
\end{equation}
and
\begin{equation}
     \big[gU\,,\,gV \big]\,\big(f(g)\big) 
                   = g\big[U,V\big] \>\big(f(g)\big)\,. 
\label{third-g}
\end{equation}
Taking eqs.~(\ref{first-g})-(\ref{third-g}) to eq.~(\ref{invol}) and noting
that $f$ is arbitrary, we finally arrive at
\begin{align}
    -\,\big[FU,FV\big]\,g & + g\, \big[U,V\big] = FWg - gW \nonumber \\
    & - \left(\, \big(FUg -gU\big)\,(F)\,\right)\>Vg
      + \left(\, \big(FVg -gV\big)\,(F)\,\right)\>Ug\,.
\label{invol-final}
\end{align}
The last two terms in the right hand side carry the action of the vector
fields $\,F(g)Ug -gU\,$ and $\,F(g)Vg -gV\,$ on $F(g)$ as a function of $g$,
the result being two linear operators that act on $V$ and $U$.

If $F$ does not depend on $g$, the action of $\,FUg -gU\,$ and $\,FVg -gV\,$ 
on $F$ is zero and eq.~(\ref{invol-final}) simplifies to
\begin{equation}
    - \big[FU\,,FV \big]\,g +  g\,\big[U\,,V\big] =  FWg - gW\/.
\label{involutive-constant}
\end{equation}

\subsection{Reduction of isometric gluing conditions to boundary conditions}

Assume that the linear map $F(g)$ is such that steps 1 and 2 are
satisfied. There is then a submanifold~$N$ whose tangent bundle $\Pi^N$ is
formed by the tangent spaces\, $T_gN=\textnormal{Im}({\cal F}+1\big)$ \,for
all $g$ in $N$.  In what follows we show that the necessary and sufficient
condition for the gluing condition~(\ref{GC-ts}) to be equivalent to a
boundary condition~(\ref{BC-sigma}) is that $F(g)$ is an isometry of the Lie
algebra metric~$\Om$.

Since\, $\pa_\tau\/x$ \,belongs to $T_gN$, there exists~$v$ in\, $T_{g}G$
\,such that\, \hbox{$\pa_\tau\/x=({\cal F}+1)v$} \,and the boundary
condition~(\ref{BC-sigma}) can be recast as
\begin{equation}
  \mG\Big(u_0\,,\,\pa_\sig X \big\vert_{\pa\Sig}\Big)
    = \om\Big(u_0\,,\,\big({\cal F} + 1\big)v\Big)
  \quad \textnormal{for all}\quad 
      u_0\in\textnormal{Im}\big({\cal F}+1\big)\,.
\label{Bfromboundary}
\end{equation}
The gluing condition~(\ref{GC-ts}) can in turn be written as
\begin{equation*}
   \big( {\cal F} + 1 \big)\, \pa_\sig X \big\vert_{\pa\Sig} 
   =  \big( {\cal F} - 1 \big)\big({\cal F}+1\big)v  \,.
\end{equation*}
This can be viewed as an equation in $\pa_\sig X\/\big\vert_{\pa\Sig}$, whose
solutions are of the form
\begin{equation}
   \pa_\sig X \big\vert_{\pa\Sig}  =  \big( {\cal F} - 1 \big)v + v_0 \,,
  \label{sol_par}
\end{equation}
with arbitrary $v_0$ in $\textnormal{Ker}\big({\cal F}+1\big)$.
Eq.~(\ref{sol_par}) implies that
\begin{equation}
   \mG\big(u_0\,,\,\pa_\sig X \big\vert_{\pa\Sig}\,\big)
       = \mG\big(u_0\,,\,({\cal F} - 1)v\big) 
       + \mG\big(u_0,v_0\big)
  \quad \textnormal{for all}\quad 
      u_0\in\textnormal{Im}\big({\cal F}+1\big)\,.
\label{Bfromgluing}
\end{equation}
Of the two terms on the right hand side, only the first one is linear in
$v$. From this and the linearity in $v$ of the boundary
condition~(\ref{Bfromboundary}), we conclude that eq.~(\ref{Bfromgluing}) is
compatible with the boundary condition~(\ref{Bfromboundary}) if and only if
the following two requisites
are met:\\[3pt]
\indent(1) $\mG\big(u_0,v_0\big) = 0$ \,for all $u_0$ in
\,$\textnormal{Im}\big({\cal F}+1\big)$ \,and all\, $v_0$ in
\,$\textnormal{Ker}\big({\cal F}+1\big)$, and \\[3pt]
\indent(2) the action of the two-form $\om$ on arbitrary\, $({\cal F}+1)u$
\,and\, $({\cal F}+1)v$ \,in\, $T_{g}N$ \,is given by
\begin{equation}
  \om\big(\,({\cal F}+1)\,u\,,\, ({\cal F}+1)\,v\,\big)
        = \mG\big(\,({\cal F}+1)\,u\,,\,({\cal F}-1)\,v\,\big)\,. 
\label{B}
\end{equation}

For eq.~(\ref{B}) to make sense, its right hand side must be
antisymmetric,
\begin{equation*}
   0 = \mG\,\big(\, ({\cal F}+1)u\,,\, ({\cal F}-1)v\,\big) 
     + \mG\,\big(\, ({\cal F}+1)v\,,\, ({\cal F}-1)u\,\big)
     = 2\, \mG\big({\cal F}u\,,{\cal F}v\big)
     - 2\,\mG\big(u\,,v\big)\,.
\end{equation*}
Since $u$ and $v$ are arbitrary in $T_gG$, the operator ${\cal F}$, defined on
$N$, acts isometrically on the whole tangent space $T_gG$,
\begin{equation}
   \mG\big({\cal F}u\,,{\cal F}v\big)
     = \mG\big(u\,,v\big)\,.
\label{isometry}
\end{equation}
This in turn implies that
\begin{equation}
   \textnormal{Im}\,( {\cal F} \pm 1)
	= \textnormal{Ker}\,( {\cal F} \pm 1)^\bot
\label{KerIm}
\end{equation}
and makes condition (1) trivial. Furthermore, given $v$ in $T_gG$, consider\,
$v'\!=v+v'_0$ \,in\, $T_gG$, with arbitrary $v'_0$ in
\,$\textnormal{Ker}({\cal F}+1)$. From eq.~(\ref{KerIm}) it follows that
\begin{equation*}
    \mG\big(\,({\cal F}+1)\,u\,,\,({\cal F}-1)\,v'\,\big) 
           =  \mG\big(\,({\cal F}+1)\,u\,,\,({\cal F}-1)\,v\,\big)\,. 
\end{equation*}
In other words, the right hand side in eq.~(\ref{B}) does not depend on the
choice of $v_0$ in (\ref{sol_par}) and the two-form $\om$ as defined by
eq.~(\ref{B}) is single valued. Finally, $\om$ exists globally on $N$ since it
is given by eq.~(\ref{B}) through its action on arbitrary vectors\, $({\cal
  F}+1)u$ \,and\, $({\cal F}+1)v$ \,in\, $T_gN$ for any $g$ in $N$.

From the definition~(\ref{calF}) of~${\cal F}$, the bi-invariance
property~(\ref{bi-invariance}) of the metric $\mG$ and eq.~(\ref{isometry}),
it is straightforward that
\begin{equation}
   \Om\big(F(g)T_A, F(g)T_B\big)=\Om(T_A,T_B)
\label{restriction}
\end{equation}
for all $T_A$ and $T_B$ in the Lie algebra. This shows that the linear map
$F(g)$ is an isometry of the Lie algebra metric $\Om$.

Note that if ${\cal F}$ is an isometry on $T_gG$, eq.~(\ref{Bfromgluing}) not
only follows from the gluing condition~(\ref{GC-ts}) but is equivalent to
it. All in all we have that the necessary and sufficient condition for the
gluing condition to have the form of a boundary condition is that $F(g)$ is an
isometry of $\Om$, the two-form $\om$ being given by eq.~(\ref{B}).  In what
follows whenever we write $F(g)$ we will be thinking of it as an isometry. In
terms of the fields\,
$t_{A\!}=g\/\big[\textnormal{Ad}_{g^{-1}}F(g)-1\big]\/T_A$ \,in
eq. ~(\ref{t_A}), the definition of $\om$ in eq~(\ref{B}) can be written as
\begin{equation}
   \om\,\big(t_A\,,\,t_B\big) 
         = \,\Omega\,\big(\/(\textnormal{Ad}_{g^{-1}}F-1)\,T_A\,,\,
         (\textnormal{Ad}_{g^{-1}}F+1)\,T_B \big) \,.
\label{B2}
\end{equation}
We remark that the analysis performed here holds for any linear map $F$,
regardless of whether it is constant or $g$-dependent.

The condition that $\om$ must satisfy for $N$ to be a D-brane is
\begin{equation}
  d\om=\mH\big\vert_N \,,
  \label{HdBinN}
\end{equation}
where the exterior derivative on the left hand side is taken with respect to
the directions in $T_gN$ and not with respect to arbitrary directions in
$T_gG$.  Condition~(\ref{HdBinN}) does not hold for every isometry $F$
defining a submanifold $N$ upon integration of the gluing
condition. Examples of this are given in Section 6 and in
ref.~\cite{HHRR}. Let us discuss some cases in which $\om$ fulfills
eq.~(\ref{HdBinN}).

For one and two-dimensional submanifolds $N$, eq.~(\ref{HdBinN}) trivially
holds and the only requirement for the existence of a D-brane for an isometry
$F$ is involutivity.  Assume now that $N$ has dimension larger than two and
that\, $F=R$ \,is a $g$-independent, $\Om$-preserving Lie algebra
automorphism. The exterior derivative of $\om$ on $N$ is a three-form whose
action on vector fields $t_1,\,t_2$ and $t_3$ in $T_gN$ is given by
\begin{align}
   d\om\,\big(t_1,\,t_2,\,t_3\big) & = t_1 \big(\,\om(t_2,t_3)\,\big) 
                     - \om\,\big(\,[t_1,t_2]\,,\,t_3\big) \nonumber\\
   & + t_2 \big(\,\om(t_3,t_1)\,\big) 
                     - \om\,\big(\,[t_2,t_3]\,,\,t_1\big) \nonumber\\
   & + t_3 \big(\,\om(t_1,t_2)\,\big)- \om\,\big(\,[t_3,t_1]\,,\,t_2\big)\,.
                    \label{ext_der}
\end{align}
Since the vector fields\, $t_A=g\,(\textnormal{Ad}_{g^{-1}}R-1)\,T_A$ \,span\,
$T_gN$, it suffices to calculate\, $d\om(t_A,t_B,t_C)$. For that, we need to
consider terms of the form\, $t_A\big(\om(t_B, t_C)\big)$ \,and\,
$\om\big([t_A,t_B],t_C\big)$.  Since $R$ does not depend on $g$,
\begin{equation*}
  t_A\,\big(\,\om(t_B,t_C)\,\big) = 
   \Omega\,\Big(\,t_A\,\big(\textnormal{Ad}_{g^{-1}}\big)\,RT_B\,,\, T_C\Big) 
    - \Omega\,\Big(\,t_A\,\big(\textnormal{Ad}_{g^{-1}}\big)\,RT_C\,,\,T_B\Big)\,.
\end{equation*}
Noting that 
\begin{equation*}
  t_A\,\big(\textnormal{Ad}_{g^{-1}}\big)\,V
   = - \big[\,\big(\textnormal{Ad}_{g^{-1}} R-1\big)\,T_A \, 
           ,\,\textnormal{Ad}_{g^{-1}}V\,\big]
\end{equation*}
for all $V$ in $\ag$, using that $R$ and $\textnormal{Ad}_{g^{-1}}R$ are Lie
algebra automorphisms and recalling that $\Omega$ is invariant, it is
straightforward to see that
\begin{equation*}
  t_A\,\big(\,\om(t_B,t_C)\,\big) =\,
      \Om\,\big(\textnormal{Ad}_{g^{-1}}\,RT_B\,,\,[\,T_C\/,T_A]\,\big)
    -\, \Om\,\big(\/T_B\,,\, \textnormal{Ad}_{g^{-1}}\,R\,[\,T_C\/,T_A]\,\big)
     \,-\,\big( B\,\leftrightarrow\,C\big)\,.
\end{equation*}
Proceeding similarly with $\om\big([t_A,t_B],t_C\big)$, one has
\begin{equation*}
  \om\,\big([t_A,t_B]\,,\,t_C\big)=\,
    \Om\,\big([\,T_A\/,T_B]\,,\,\textnormal{Ad}_{g^{-1}}\,RT_C\big)
    -\,\Om\,\big(\/\textnormal{Ad}_{g^{-1}}\,R\,[\,T_A\/,T_B]\,,\,T_C\big) \,.
\end{equation*}
Upon substitution in eq.~(\ref{ext_der}), this gives 
\begin{align}
  d\om(t_A,t_B,t_C) = \Omega\,\Big(\big[\,(\textnormal{Ad}_{g^{-1}}R-1)\,T_A\,,
    \, (\textnormal{Ad}_{g^{-1}}R-1)\,T_B\,\big]\,,
    \, (\textnormal{Ad}_{g^{-1}}R-1)\,T_C\Big)\,.
\label{dom}
\end{align}
On the other hand, from eq.~(\ref{H}) it trivially follows that the right hand
side in~(\ref{dom}) is equal to\, $\mH(t_A,t_B,t_C)$. Hence, for any constant
isometry that is also a Lie algebra automorphism, $d\om=\mH\big\vert_N$.  It
is clear that if $F$ depends on $g$ and/or is not a Lie algebra automorphism,
this proof does not stand. In these cases, condition~(\ref{HdBinN}) can always
be checked by hand. See ref.~\cite{HHRR} for examples.

We end this section by remarking that we have not assumed at any stage that\,
$T_gG=\Pi_g \oplus \Pi_g^\bot$, thus generalizing previous
approaches~\cite{Stanciu-note} that, under such an assumption, define~$\om$
for $F$ a constant Lie algebra automorphism. In this regard, it is worth
noting that $T_gG=\Pi_g\oplus \Pi_g^\bot$ holds for Lie groups with Euclidean
signature metric $\mG_{\!\m\n}$. However, if $\mG_{\!\m\n}$ is Lorentzian, it
may occur that, among the vector fields defining the distribution $\Pi^M$, one
of them is null and orthogonal to all the others.  If this is the case, the
induced metric on the D-brane is degenerate and the tangent space $T_gG$
cannot be written as a direct sum of $\Pi_g$ and $\Pi_g^\bot$. In Appendix B
an explicit construction of such null vector fields in terms of the
eigenvectors of ${\cal F}$ is presented, and in ref.~\cite{HHRR} a family of
degenerate D2-branes for the Nappi-Witten~\cite{Nappi-Witten} model is found.

\section{Limitations of the gluing condition approach}

In the previous section we have shown that every gluing condition with $F$ an
isometry takes the form of a boundary condition with a two-form $\om$ defined
by eq.~(\ref{B}). It may, however, occur that a boundary condition describing
a D-brane cannot be written as a gluing condition. In this section we tackle
this problem and show that every boundary condition with two-form $\om$
defining a D-brane $N$ can be written as a gluing condition if and only if\,
$\textnormal{det}\,(\mG\vert_N-\om)\neq\/0$.

Let us then consider a D-brane $N$ with tangent space $T_gN$ specified by the
boundary condition~(\ref{BC-sigma}), the two-form $\om$ acting on $T_gN$. It
is most convenient for our purposes to write the boundary condition as
\begin{equation}
  \mG\Big(\dl\/X\,,\,\pa_\sig X \big\vert_{\pa\Sig}\Big)
    = \om\Big(\dl\/X\,,\,\pa_\tau X \big\vert_{\pa\Sig}\Big)
  \quad \textnormal{for all}\quad 
      \dl\/X\in\/T_gN\,,
\label{bounalt}
\end{equation}
with $\pa_\sig X \big\vert_{\pa\Sig}$ in $T_gG$ and $\pa_\tau X
\big\vert_{\pa\Sig}$ in $T_gN$. We now define a map\, ${\cal K}\!: T_gN\to
T_gG/(T_gN^\bot)$ \,whose action on $w$ in $T_gN$ is given by
\begin{equation}
   \mG(z\,,{\cal K}w) =\,\frac{1}{2}\>\big[\, \mG(z,w) - \om(z,w)\,\big]
   \quad \textnormal{for all}\quad z\in\/T_gN\,.
\label{mapK}
\end{equation}
The map ${\cal K}$ is trivially linear and takes values in the quotient
$T_gG/(T_gN^\bot)$. To see the latter, assume that $y$ in\, $T_gG$ \,is such
that\, $\mG(z,y)=\mG(z,{\cal K}w)$ \,for all $z$ in $T_gN$. It follows that
$\mG(z,y-{\cal K}w)=0$, which in turn implies that $y-{\cal K}w$ is in
$T_gN^\bot$.

Furthermore, ${\cal K}$ is injective if and only if\,
$\textnormal{det}\,(\mG\vert_N-\om)\neq\/0$.  Indeed, for\, $w'\neq\/w$ \,in\,
$T_gN$ \,such that ${\cal K}w^{\,\prime}\!={\cal K}w$, we have, according to
eq.~(\ref{mapK}), that
\begin{equation}
   \mG(z,w'-w)=\om(z,w'-w)\quad  \textnormal{for all}\quad z\in\/T_gN\,.
\label{zww}
\end{equation}
A vector\, $w^{\,\prime\!}-w\neq\/0$ \,satisfying this condition exists if and
only if\, $\textnormal{det}\,(\mG\vert_N-\om)=0$, which proves the statement.
Actually, since\, $\textnormal{dim}\/(T_gN)+\textnormal{dim}(T_gN^\bot)
=\textnormal{dim}(T_gG)$, the map ${\cal K}$ is bijective if it is injective.

For $\textnormal{det}\,(\mG\vert_N-\om)\neq\/0$, the inverse map\, ${\cal
  K}^{-1}\!: T_gG/(T_gN^\bot)\to T_gN$ hence exists and is bijective. From
${\cal K}^{-1}$ we define a linear map ${\cal G}\!: T_gG\to T_gN$ whose action
on an arbitrary element $v$ in $T_gG$ is given by ${\cal G}v ={\cal
  K}^{-1}(v+T_gN^\bot)$. Writing ${\cal G}$ as ${\cal G}={\cal F}+1$, it is
straightforward to check that ${\cal F}$ satisfies\,
$T_gN=\textnormal{Im}({\cal F}+1)$ \,and
  \begin{equation}
  \om\big(\,({\cal F}+1)\,u\,,\,({\cal F}+1)\,v\,\big) 
           =  \mG\big(\,({\cal F}+1)\,u\,,\,({\cal F}-1)\,v\,\big)\,
\label{ombis}
\end{equation}
for arbitrary $u$ and $v$ in $T_gG$. Proceeding along the same lines as in
Subsection 3.3, one can see that the isometric character of $\cal F$ follows
from the antisymmetric property of $\om$.

Since $\pa_\tau X\big\vert_{\pa\Sig}$ in eq.~(\ref{bounalt}) belongs to
$T_gN$, it can be written as\, $\pa_\tau X\big\vert_{\pa\Sig}=({\cal F}+1)v$
\,for some $v$ in $T_gG$. Upon noting~(\ref{ombis}), the boundary
condition~(\ref{bounalt}) takes the form 
\begin{equation*}
  \mG\Big(\dl\/X\,,\,\pa_\sig X \big\vert_{\pa\Sig}\Big)
    = \mG\Big(\dl\/X\,,\,({\cal F}-1)v\Big)
  \quad \textnormal{for all}\quad 
      \dl\/X\in\/T_gN\,.
\end{equation*}
This is equivalent to 
\begin{equation*}
\pa_\sig X \big\vert_{\pa\Sig}  
  =  \big( {\cal F} - 1 \big)v + v_0\,,
\end{equation*} 
with arbitrary $v_0$ in\, $T_gN^\bot\!=\textnormal{Ker}\big({\cal F}+1\big)$.
Acting with ${\cal F}+1$ on both sides of this equation we finally have 
\begin{equation*}
   \big( {\cal F} + 1 \big)\, \pa_\sig X \big\vert_{\pa\Sig} 
   =  \big( {\cal F} - 1 \big)\big({\cal F}+1\big)v 
   = \big( {\cal F} - 1 \big)\pa_\tau X\big\vert_{\pa\Sig}\,,
\end{equation*}
which is nothing but the gluing condition~(\ref{GC-ts}) written in terms of
world sheet coordinates~$\tau$ and~$\sig$.

Let us go back to eq.~(\ref{zww}). If there exists\, $w^{\,\prime\!}-w\neq\/0$
\,in\, $T_gN$ \,such that the equation holds and $z$ is taken equal to
$w^{\,\prime\!}-w$, the right hand side of eq.~(\ref{zww}) vanishes and it
follows that $w^{\,\prime\!}-w$ is a null vector.  Since Euclidean D-branes do
not have null vectors, such a $w^{\,\prime\!}-w\neq\/0$ does not exist and\,
$\textnormal{det}\,(\mG\vert_N-\om)\neq\/0$. The analysis of D-branes based on
the gluing condition~(\ref{GC-general}) then provides all Euclidean D-branes
described by boundary conditions but may miss some Lorentzian or metrically
degenerate D-branes for which $\textnormal{det}\,(\mG\vert_N-\om)=\/0$.

\section{An application: D-branes from global isometries}

From the analysis in Subsection 3.2 it is convenient to distinguish two
cases. The first one assumes that $F$ does not depend on $g$. We call such
isometries constant or global and will be treated in this section. The
second case accounts for $g$-dependent isometries $F(g)$. We call them local
or nonconstant; some examples will be considered in Section 5.

If $F$ is a global (or constant) isometry solving involutivity,
eq.~(\ref{involutive-constant}) holds. Frobenius theorem ensures that\,
$\Pi_g=g(\textnormal{Ad}_{g^{-1}}F-1)\ag$ \,is the tangent space to a
submanifold $N$ of $\mG$ but it does not identify $N$. This problem we address
next.

Consider the vector field\, $t_V(g)=g(\textnormal{Ad}_{g^{-1}}F-1)V$, with $V$
in $\ag$, and let $g_0$ be a group element. By definition, the integral curve
$\ga_{t_V}(s;g_0)$ of $t_V$ that goes through $g_0$ is the solution to
\begin{equation*}
  \ga_{t_V}(0;g_0)=g_0 \qquad  
  \frac{d}{ds}\ga_{t_V}(s;g_0)=t_V\big(\ga_{t_V}(s;g_0)\big) \,,
\end{equation*}
where $s$ is a real parameter along the curve. Simple inspection shows that
the solution is
\begin{equation}
   \ga_{t_V}(s;g_0)= e^{\,sFV}g_0\,e^{-sV} \,.
\label{int-curve}
\end{equation}
The set $N_{\!g_0}$ of all points connected to $g_0$ by integral curves of
vector fields $t_V$, with $V$ arbitrary, can always be written as
\begin{equation}
   N_{g_0}= \big\{ e^{\,FV}g_0\,e^{-V}\!\!: ~V\!\in\ag \big\} \,.
\label{N}
\end{equation}
The only candidate for a D-brane containing $g_0$ is then
$N_{\!g_0}$. According to Section 3, however, the fact that $N_{\!g_0}$
contains the integral curves of all the fields $t_V$ that go through $g_0$ is
not enough to conclude that $N_{g_0}$ is a D-brane. For this to be the case,\,
\hbox{$\Pi^{N_{g_0}}$} \,must be an involutive distribution. In summary,
D-branes for constant $F$, \,if they exist, have the form of $N_{\!g_0}$ in
(\ref{N}).

\subsection{Automorphisms and twined conjugacy classes as D-branes}

Take $F=R$ with $R$ a Lie algebra automorphism compatible with
condition~(\ref{restriction}). Automorphisms of this type form a group,
denoted by $\textnormal{\sl Aut}_\Om(\ag)$. Being $R$ an automorphism, it
satisfies
\begin{equation} 
    R[\,U,V] = [\,RU,RV] 
\label{automorphism}
\end{equation}
for all~$U$ and~$V$ in~$\ag$. For any such $F$, the involutivity
equation~(\ref{involutive-constant}) is solved by \hbox{$\,W=[V,U]\,$} and the
manifold~$N_{g_0}$ is the \hbox{$R$-twined} conjugacy class $\,{\cal
  C}(R,g_0)\,$ of~$\,g_0$,
\begin{equation*}
  N_{g_0}={\cal C}(R,g_0) = \big\{ e^{\,RV}g_0\,e^{-V}\!\!: 
  ~V\!\in\ag \big\} \,.
\end{equation*}
In Appendix C it is shown that the dimension of
$\,\Pi_g=g(\textnormal{Ad}_{g^{-1}}F-1)\ag\,$ is constant for all $g$ in
${\cal C}(R,g_0)$.  Furthermore, as proved at the end of Section 3, the
two-form $\om$ given in~(\ref{B}) satisfies
$H\big\vert_{N_{g_0}}=d\om$. Hence, ${\cal C}(R,g_0)$ is a
D-brane~\cite{Alekseev-Schomerus,Stanciu-manifolds,Stanciu-note,FS-more}.
  
Note that for $R=1$, the manifold $N_{g_0}$ is a conventional conjugacy class
$\,{\cal C}(1,g_0)$~\cite{Alekseev-Schomerus}. Consider now $\,R\neq 1$ and
assume that $R$ is an inner automorphism. By definition, it exists an $h$ in
$G$ such that \hbox{$\,RV\!={\rm Ad}_hV$}\, for all $V$ in $\ag$. Since
$\>\textnormal{Ad}_{h_1}\textnormal{Ad}_{h_2}\!=\textnormal{Ad}_{h_1h_2}$,
inner automorphisms form a subgroup $\,\textnormal{\sl
  Inn}_\Om(\ag)$. Automorphisms which are not inner are called outer and form
the equivalence classes of the quotient $\textnormal{\sl
  Aut}_\Om(\ag)/\textnormal{\sl Inn}_\Om(\ag)$. Any automorphism $R$ can
therefore be written as $R=R_1\tilde{R}_2$, with $R_1$ inner and $\tilde{R}_2$
of the same type as $R$. Consider the $(R_1\tilde{R}_2)$-twined conjugacy
class of $g_0$. Using that $R_1\!=\textnormal{Ad}_h$ for some $h$ in~$G$, and
recalling that $e^{\,t\textnormal{Ad}_r U}=\textnormal{Ad}_r e^{tU}$ for all
$r$ in $G$ and all $U$ in $\ag$, it follows that
\begin{equation*}
  {\cal C}(R_1\tilde{R}_2,g_0) 
      = h\,{\cal C}(\tilde{R_2},h^{-1}g_0) 
\end{equation*}
for some $h$ in $G$. If $R$ is inner, so that $\tilde{R}_2=1$, the D-branes
are the left translates by $h$ of the conventional conjugacy
classes~\cite{Stanciu-manifolds}. For $R$ outer, the D-branes are the
translates of $\tilde{R}_2$-twined conjugacy
classes~\cite{Birke,Stanciu-manifolds}.

\subsection{D-branes for semisimple Lie algebras}

Involutivity ~for isometries that are not Lie algebra automorphisms is
more complicated. It has been suggested~\cite{Stanciu-3,Stanciu-manifolds,
  Stanciu-Tseytlin} that isometries of the form $F=-R$, with $R$ a Lie algebra
automorphism, define D-branes. In the sequel we investigate this issue and
reach an answer in the negative.

Consider eq.~(\ref{involutive-constant}) and make the change $\,2Y=[U,V]-W$.
Involutivity requires that, for all $U$ and $V$ in $\ag$
and for all $g$, there must exist $Y$ in $\ag$ such that
\begin{equation*}
  g\,[U,V]=g\big(\textnormal{Ad}_{g^{-1}}R+1\big)Y \,.
\end{equation*}
After multiplying from the left with $g^{-1}$, this becomes 
\begin{equation}
  [\ag,\ag]\subset \big(\textnormal{Ad}_{g^{-1}}R+1\big)\ag \,.
\label{anti}
\end{equation}

We restrict ourselves to semisimple Lie algebras $\ag$. Concerning this
restriction, we make two comments. The first one is that if for the invariant
metric $\Om$ one takes a Killing form and $R$ is an arbitrary Lie algebra
automorphism, $\pm\/R$ are isometries\footnote{Note in this regard that a
  non-simple semisimple Lie algebra may have invariant metrics which are not
  proportional to its Killing form, in which case a Lie algebra automorphism
  need not be an isometry. The case of simple Lie algebras is more restrictive
  since all invariant metrics are proportional to the Killing form.}.  The
second observation is that two of the most relevant semisimple Lie algebras in
string theory are $\as\al(2,{\bf R})$ and $\as\au(2)$, all whose isometries
are either of the form $F=R$ or $F=-R$.  As already mentioned, $F=R$ define
D-branes. Consideration of\, $F=-\!R$ \,then completes the search of D-branes
for constant isometries for $\as\al(2,{\bf R})$ and $\as\au(2)$.

Let $\ag$ be a semisimple Lie algebra of dimension ${\tt d}$. Being
semisimple, $[\ag,\ag]=\ag$ and the involutivity requirement~(\ref{anti})
reads
\begin{equation*}
  \ag = \big(\textnormal{Ad}_{g^{-1}}R+1\big)\ag  \,.
\end{equation*}
Let $D_{\tt d}$ denote the set of group elements where this condition holds.
That is, $g$ belongs to $D_{\tt d}$ if
$(\textnormal{Ad}_{g^{-1}}R+1)V\!\neq\!\/0$\, for all $V\neq 0$ in $\ag$.  The
tangent plane~(\ref{Tg}) at all $g$ in $D_{\tt d}$ is then $\Pi_g= T_gG$. The
distribution $\Pi^{D_{\tt d}}$ is trivially involutive and is the tangent
bundle to $D_{\tt d}$ itself. The only D-brane candidate provided by
$F=\!-R$ is hence $D_{\tt d}$. According to our discussion in Section 3, for
$D_{\tt d}$ to be a D-brane, the integral curves of the vector fields
$t_V(g)=g(\textnormal{Ad}_{g^{-1}}R+1)V$ should be contained in~$D_{\tt
  d}$. In the remaining of this section we show that this is not the case,
thus implying that $F=\!-R$ does not define a D-brane.

The proof consists in {\bf (i)} finding group elements $g$ outside $D_{\tt d}$, and
{\bf (ii)} showing that these $g$ are connected to elements in $D_{\tt
  d}$ by integral curves of the vector fields $t_V(g)$. 

{\bf Proof of  (i)}. A group element $g$ is not in $D_{\tt d}$ if there exists a
nonzero $V$ in $\ag$ such that\,
$\big(\textnormal{Ad}_{g^{-1}}R+1\big)V=0$. \,Let us call $ D^-_{\tt d}$ to
the set formed by such $g$,
\begin{equation}
   D^-_{\tt d} = G-D_{\tt d}=\big\{ g\in\/G \!:
      ~\textnormal{Ker}\big(\textnormal{Ad}_{g^{-1}}R+1\big)\neq0\big\}\,.
\label{Q}
\end{equation}
The group $\textnormal{Aut}(\ag)$ has in general several connected components,
the component containing the identity being the normal subgroup
$\textnormal{\sl Inn}(\ag)$, and the quotient
$\textnormal{\sl Aut}(\ag)/\textnormal{\sl Inn}(\ag)$ being a finite group. It follows
that in every component, and in particular in that containing $R$, there is
then an automorphism $S$ such that ${S}^n\!=1$ for an integer $n$. This
implies that the eigenvalues of $S$ can only be $n$-th roots of~1. Furthermore,
if a root $e^{-\imu\th}$ is an eigenvalue with multiplicity $m$, so is
$e^{\imu\th}$.  Since $S$ and $R$ are in the same component, they are related
by an inner automorphism, meaning that there is an $h$ in $G$ such that\,
$R=\textnormal{Ad}_h S$. We now distinguish three cases:
\begin{itemize}
\vspace{-6pt}
\item $S$ has an eigenvalue $-1$. For $g=h$, the operator
  $\textnormal{Ad}_{g^{-1}}R$ has then an eigenvalue $-1$ and $g$ is in~$D^-_{\tt
    d}$.
\vspace{-6pt}
\item $S$ does not have an eigenvalue $-1$ but has two complex conjugate
  eigenvalues $e^{\mp\imu\theta}$ with eigenvectors\,
  $Z_1\pm\/\imu\/Z_2$, 
\begin{equation*}
    S (Z_1\pm\/\imu Z_2)=e^{\mp\imu\theta}(Z_1\pm\imu Z_2)\,.
\end{equation*}
Using that $\ag$ is semisimple and that $S$ is an automorphism, it is
straightforward to show that $X=[Z_1,Z_2]$ is an eigenvector of $S$ with
eigenvalue $+1$ and that there exists $a$ real such that\, $[X,Z_1]= aZ_2$
\,and\, $[X,Z_2]=-aZ_1$. The constant $a$ can be eliminated by redefining
$Z_1,\,Z_2$ and $X$, so that 
\begin{equation*}
  [Z_1,Z_2]=X \qquad [X,Z_1]= Z_2 \qquad    [X,Z_2]=-Z_1 \,.
\end{equation*}
Take now\, $W\!=\frac{\pi}{\sqrt{2}}(Z_1\!-Z_2)$ \,and consider\,
$g=he^{-W}$. It follows after some algebra that the automorphism\,
$\textnormal{Ad}_{g^{-1}}R= \textnormal{Ad}_{e^{W}}S$ \,has two eigenvectors
with eigenvalue $-1$,
\begin{alignat*}{3}
    \textnormal{Ad}_{g^{-1}}R\,X &= - X &\\
    \textnormal{Ad}_{g^{-1}}R\,\big(\cos \theta Z_1+Z_2-\sin\theta Z_2 \big)
         &= - \big(\cos \theta Z_1+Z_2-\sin\theta Z_2 \big)\,.&
\end{alignat*}
Hence $g=he^{-W}$ belongs to $D^-_{\tt d}$. 
\vspace{-6pt}
\item $S$ only has eigenvalues $+1$. In this case, $S$ is the identity
  automorphism and $R$ is inner. For $\ag$ semisimple, it is always possible
  to take $X$ in its Cartan subalgebra and $Z_1$ and $Z_2$ in $\ag$ such that
\begin{equation*}
  [Z_1,Z_2]=  X \qquad  [X,Z_1]= Z_1 \qquad    [X,Z_2]=-Z_2 \,\,.
\end{equation*}
It is then very simple to check that\, $\textnormal{Ad}_{g^{-1}}R$,
where $g$ is taken as $g=he^{-W}$ with $W=\sqrt{2}\pi(Z_1\!-Z_2)$, has two
eigenvectors with eigenvalue $-1$,
\begin{alignat*}{3}
    \textnormal{Ad}_{g^{-1}}R\,X &= - X &\\
    \textnormal{Ad}_{g^{-1}}R\,\big(Z_1+Z_2\big)
         &= - \big(Z_1+Z_2)\,.&
\end{alignat*}
So also in this case $D^-_{\tt d}$ is not empty.
\end{itemize}
\vspace{-6pt} As shown in Appendix C, the spectrum of
$\textnormal{Ad}_{g^{-1}}R$ is invariant under $R$-twined conjugation. Hence,
if $g$ is in $D^-_{\tt d}$, the whole $R$-twined conjugacy class\, ${\cal
  C}(R,g)$ \,is in $D^-_{\tt d}$. As a result, $D^-_{\tt d}$ is a
union of \hbox{$R$-twined} conjugacy classes. It is clear that $D^-_{\tt d}$
  has dimension less than ${\tt d}$.

{\bf Proof of (ii)}. The tangent space $T_{g_0}G=g_0\ag$ at a $g_0$ in
$D^-_{\tt d}$ is most conveniently written as
\begin{equation}
   T_{g_0}\mG= g_0\,(\textnormal{Ad}_{g_0^{-1}}R-1)\,\ag 
     ~\,\cup~\, g_0\,(\textnormal{Ad}_{g_0^{-1}}R+1)\,\ag\,.
\end{equation}
Since the \hbox{$R$-twined} conjugacy class\, ${\cal C}(R,g_0)$ is contained
in $D^-_{\tt d}$ and the fields\, $g(\textnormal{Ad}_{g^{-1}}R-1)\ag$ generate
motions inside ${\cal C}(R,g_0)$, there must be at least one vector field\,
$t_V(g)=g\,(\textnormal{Ad}_{g^{-1}}R+1)V$ \,whose integral curve goes from
$D^-_{\tt d}$ to $D_{\tt d}$.  Such a curve connects points in the D-brane
with points outside the D-brane.  We thus conclude that $D_{\tt d}$ cannot be
a D-brane.

This proves that there are no D-branes for a semisimple Lie algebra $\ag$ and
$F=-R$, with~$R$ a constant automorphism. This result contrasts with previous
studies on the subject~\cite{Stanciu-3}. If the requirement that $D_{\tt d}$
contain the integral curves of all the fields $t_V$ were relaxed, $D_{\tt d}$
would be a D-brane of dimension ${\tt d}$, provided it exists a suitable
two-form $\om$. This D-brane would not be filling, since $D^-_{\tt d}$ is not
empty. Furthermore, it would exclude allowed motions for the string endpoints,
thus contradicting the definition of D-brane.

\section{Some considerations on D-branes for local isometries}

For local isometries $F(g)$, involutivity takes the form~(\ref{invol-final}).
Given a local isometry $F(g)$, it is always possible to construct a new
isometry
\begin{equation}
   F(g)\to F^{\,\prime}(g) =\textnormal{Ad}_g\,F^{-1}(g)\,\textnormal{Ad}_g\,.
\label{equiv}
\end{equation}
It is very easy to convince oneself that, at any point $g$ in $G$, both $F$
and $F^{\,\prime}$ define the same tangent space\,
\hbox{$\Pi_{g\!}=g\,(\textnormal{Ad}_{g^{-1}}F-1)\,\ag$}. They thus define the
same distribution. Furthermore, it is straightforward to check that
$F^{\,\prime}$ satisfies the involutivity condition~(\ref{invol-final}) if and
only if $F$ does.  Assume that this is the case, so that they define the same
submanifold $N$ of $G$.

The gluing conditions~(\ref{GC-general}) for $F$ and $F^{\,\prime}$ read
\begin{alignat}{4}
   F\! &:&({\cal F}-1)\,\pa_\tau\/ X\big\vert_{\pa\Sig} 
               &= ({\cal F}+1)\,\pa_\sig\/ X\big\vert_{\pa\Sig}  
    \label{Siuno} \\[3pt]
   F^{\,\prime}\! &:&\quad  ({\cal F}^{\,\prime}-1)\,\pa_\tau\/ X\big\vert_{\pa\Sig} 
             &= ({\cal F}^{\,\prime}+1)\,\pa_\sig\/ X\big\vert_{\pa\Sig} \,,  
    \label{Siotro}
\end{alignat}
where the matrices ${\cal F}$ and ${\cal F}^{\,\prime}$ are given by\, ${\cal
  F}\!=\!-\,\bar{e}^{\,-1} F e$ \,and\, ${\cal
  F}^{\,\prime}\!=\!-\,\bar{e}^{\,-1} F^{\,\prime} e$.  Noting that ${\cal
  F}^{\,\prime}={\cal F}^{\,-1}$, eq.~(\ref{Siotro}) can be written, after
multiplication from the left with ${\cal F}$, as
\begin{equation}
     F^{\,\prime}\! :\quad  -\,({\cal F}-1)\,\pa_\tau\/ X\big\vert_{\pa\Sig} 
               =  ({\cal F}+1)\,\pa_\sig\/ X\big\vert_{\pa\Sig}\,. 
\label{Sidos}
\end{equation}
The gluing condition~(\ref{Sidos}) for $F^{\,\prime}$ has a relative negative
sign as compared to the gluing condition~(\ref{Siuno}) for $F$.  This sign has
important consequences for the recasting of the corresponding gluing
conditions as boundary conditions.  Indeed, the two-forms $\om$ and $\om^{\,\prime}$
associated to $F$ and $F^{\,\prime}$ are related by
$\om^{\,\prime}\!=\!-\,\om$, so the conditions\, $d\om=\mH\big\vert_N$ \,and\,
$d\om^{\,\prime}=\mH\big\vert_N$ cannot generally hold simultaneously. Let us
see an example.

{\bf\emph{Example: Filling D-brane}}. In this case, the sigma model boundary
conditions~(\ref{BC-sigma}) become
\begin{equation}
    \om_{\!\m\n}\,\pa_\tau X^\n \,\Big\vert_{\pa\Sig} 
              = \mG_{\m\n}\,\pa_\sig X^\n\Big\vert_{\pa\Sig} \,.
\label{BC-filling}
\end{equation}
Assume that the D-brane is defined by an isometry $F$. This requires in
particular that the gluing condition~(\ref{Siuno}) can be written as
in~(\ref{BC-filling}), with $\om$ such that $d\om=\mH$. See ref.~\cite{HHRR}
for some examples.  The gluing condition~(\ref{Sidos}) can then be written in
the form~(\ref{BC-filling}), but needs $\om^{\,\prime}=-\om$, and
$d\om^{\,\prime}\neq\mH$. The isometry $F^{\,\prime}$ hence does not define a
D-brane.

We close this section by further illustrating that an integrable gluing
condition by itself does not define a D-brane.  Consider
$F(g)=\!-\textnormal{Ad}_g$.  The tangent plane~(\ref{Tg}) at all $g$ in $G$ is
$\Pi_g=T_g\/G$.  The isometry $F$ defines trivially an involutive
distribution, the submanifold $N$ being the whole group $G$. Since ${\cal
  F}=1$, the gluing condition~(\ref{Siuno}) becomes
$\pa_\sig\/X^\m\big\vert_{\pa\Sig}=0$. This, in turn, cannot be understood as
a sigma model boundary condition, since it requires $\om=0$ on the whole group
manifold and does not account for a nontrivial~$\mH$.

\section{Conclusion}

Given a WZW model with real Lie group $G$, Lie algebra $\ag$ and invariant Lie
algebra metric $\Om$, we have shown that a linear map $F(g)$ acting on $\ag$
defines a D-brane if the following conditions hold:
\begin{itemize}%
\vspace{-\itemsep}\vspace{-\partopsep}%
\item[(i)] $F(g)$ is an isometry of $\Om$.\vspace{-\itemsep}
\item[(ii)] The vector fields $t_A = FT_Ag-gT_A=t^\m{\!}_A\pa_\m$ defined by
  $F(g)$ span a distribution. That is, the matrix formed by the coefficients
  $t^\m{\!}_A$ has constant rank on a submanifold $N$ of the group
  manifold. If this is the case and the rank is $p+1$, there are $p+1$
  linearly independent vector fields $k_i$ that are linear combinations\,
  $k_i=c_{iA}\,t_A$ \,of the fields $t_A$. \vspace{-\itemsep}%
\item[(iii)] The integral curves of the fields $k_i$ are contained in $N$.
  \vspace{-\itemsep}%
\item[(iv)] The fields $k_i$ are involutive in $N$.
\vspace{-\itemsep}%
\item[(v)] The two-form\, $\om$ globally defined on $N$ by its action\,
  $\om(k_i,k_j)$ \,on the fields\, $k_i=c_{iA}\,t_A$ \,through\,
  $\om(t_A,t_B) =\! \Om\,\big(\,\textnormal{Ad}_{g^{-1}} FT_A -T_A\,,\,
  \textnormal{Ad}_{g^{-1}} FT_B +T_B\,\big)$ satisfies $d\om=H\big\vert_N$.
  \vspace{-\itemsep}\vspace{-\partopsep}%
\end{itemize}

The conditions above account for both metrically nondegenerate and degenerate
D-branes. They are met by $F$ any constant $\Om$-preserving Lie algebra
automorphism $R$, so the well known result~\cite{Alekseev-Schomerus,
  Stanciu-3, Stanciu-manifolds, Stanciu-note} that the $R$-twined conjugacy
classes of the group $G$ are D-branes extends 
to metrically degenerate classes.

WZW models based on semisimple Lie algebras are of particular interest in
string theory, two of the most studied models being $\as\au(2)$ and
$\as\al(2,\textnormal{\bf R})$. It had been claimed that constant $F=-R$ could
provide D-branes for such models. This has been disproved in this paper, since
condition (iii) above fails.

For more general scenarios, (ii)-(v) must be checked for any given isometry
$F$. This is however straightforward.  In ref.~\cite{HHRR} the Nappi-Witten
model~\cite{Nappi-Witten} is considered and several families of D-branes for
$g$-dependent isometries $F(g)$ are found, some have Euclidean signature, some
have Lorentzian and some are metrically degenerate. It would be
  interesting to study if D-branes defined by $g$-dependent isometries have a
  translate in the algebraic framework, since normal ordering ambiguities may
  occur.
Our interest in this paper has been the geometric description of
  D-branes in WZW string backgrounds taking as starting point a gluing
  condition $\mJ_+=F\mJ_-$ that matches the chiral currents at the world sheet
  boundary. It remains an open problem to study if the geometric approach
  presented here describes D-branes for which a full set of gluing
  conditions have not been found, the so-called permutation
  D-branes~\cite{Fredenhagen-Quella} among them.

\section*{Appendix A. Alternative derivation of eq.~(\ref{Pi-g})}
\renewcommand{\theequation}{A.\arabic{equation}}

Here we present an alternative derivation of eq.~(\ref{Pi-g}). The idea is to
solve the gluing condition~(\ref{GC-ts}) for $\pa_\tau x^\m$ in terms of the
eigenvectors of the matrix ${\cal F}$.

The (generalized) eigenvectors of the matrix ${\cal F}$ form a basis of
linearly independent vectors. An eigenvalue $\la$ with algebraic multiplicity
$a_\la$ and geometric multiplicity $m_\la$ has $i=1,\ldots,m_\la$
eigenvectors\, ${v}_{(\la\,,\,i\,,\,1)}$ \,and\, $a_\la-m_\la$ \,generalized
eigenvectors that can be organized in $m_\la$ chains
\begin{equation} 
  \big({\cal F} - \la\big)\, {v}_{(\,\la\,,\,i\,,\,1)}=0 ~~~.\>.\>.~~~ 
  \big({\cal F} - \la\big)\, {v}_{(\la\,,\,i\,,\,\ell_i)} 
            ={v}_{(\la\,,\,i\,,\,\ell_i-1)}   
  \qquad~~ \ell_i=2,\,\ldots,\, L_i\,.
\label{chain}
\end{equation}
The index\, $\ell_{i\!}=1,\ldots,L_i$ \,labels the members of the chain
$(\la,i)$. Every chain is headed by an eigenvector
$\,{v}_{(\la\,,\,i\,,\,1)}\,$ and terminates in a highest-$\ell_i$ generalized
eigenvector ${v}_{\la\,,\,i\,,\,L_i}$. Consider two arbitrary (generalized)
eigenvectors $\,{v}_{(\la\,,\,i\,,\,\ell_i)}\,$ and
$\,{v}_{(\mu\,,\,j\,,\,m_j)}\,$ relative to the eigenvalues $\la$ and
$\mu$. 

Since the (generalized) eigenvectors $\,\{v_{(\la\,,\,i\,,\,\ell_i)}\}\,$ are
linearly independent, $\,\pa_\tau X\big\vert_{\pa\Sig}\,$ and $\,\pa_\sig
X\big\vert_{\pa\Sig}\,$ are linear combinations
\begin{equation*}
  \pa_\tau X\big\vert_{\pa\Sig} = \sum_{\la,i,\ell_i}\> 
      \a_{(\la\,,\,i\,,\,\ell_i)}\;v_{(\la\,,\,i\,,\,\ell_i)}\,  \qquad
   \pa_\sig X\big\vert_{\pa\Sig} = \sum_{\la,i,\ell_i}\> 
      \b_{(\la\,,\,i\,,\,\ell_i)}\,v_{(\la\,,\,i\,,\,\ell_i)}\,,
\end{equation*}
with coefficients $\,\a_{(\la\,,\,i\,,\,\ell_i)}\,$ and
$\,\b_{(\la\,,\,i\,,\,\ell_i)}$. Upon substitution in eq.~(\ref{GC-ts}), the
following set of equations follows for every chain $(\la,i)$
\begin{align}
    \a_{(\la\,,\,i\,,\,\ell_i)} + (\la-1)\; \a_{(\la\,,\,i\,,\,\ell_i-1)} 
       & =  \b_{(\la\,,\,i\,,\,\ell_i)} + (\la+1)\; \b_{(\la\,,\,i\,,\,\ell_i-1)} 
            \qquad \ell_i=2,\dots,L_i \label{tan-1} \\[3pt]
    (\la-1)\; \a_{(\la\,,\,i\,,\,L_i)} & = (\la+1)\; \b_{(\la\,,\,i\,,\,L_i)}  \,.  
   \label{tan-2}
\end{align}
We must solve eqs.~(\ref{tan-1})-(\ref{tan-2}) for $\,\a_{(\la,i,\ell_i)\,}$
in terms of $\,\b_{(\la,i,\ell_i)\,}$. To this end, we consider the cases
$\,\la=-1$, $\,\la=1\,$ and $\,\la\neq\pm 1\,$ separately.

{\leftskip=\parindent 
  \noindent $\bullet$ Assume that ${\cal F}$ has a chain\,
  $\{v_{(-1,i,\ell_i)}\}$ \, relative to the eigenvalue
  $\la=\!-\,1$. Eq.~(\ref{tan-2}) implies\, \hbox{$\a_{(-1,i,L_i)}\!=0\,$}, so
  the vector $\,v_{(-1,i,L_i)}\,$ does not occur in
  $\,\pa_\tau{}x$. Eq.~(\ref{tan-1}) in turn implies that there are infinitely
  many solutions for\, $\a_{(-1,i,1)}\ldots\a_{(-1,i,L_i-1)\,}$; \,one for
  every choice of $\,\b_{(-1,i,1)\,}\ldots \b_{(-1,i,L_i,)\,}$. The
  (generalized) eigenvectors $\,v_{(-1,i,1)} \ldots\/v_{(-1,i,L_i-1)}\,$ then
  occur in $\pa_\tau\/x$.

  \noindent $\bullet$ Look next at a chain\, $\{v_{(1,i,\ell_i)}\}$ \,with
  eigenvalue $\,\la=1$. Eq.~(\ref{tan-2}) now requires $\,\b_{(1,i,L_i)}\!=0$
  and leaves $\,\a_{(1,i,L_i)}$ arbitrary. This and eq.~(\ref{tan-1}) give
  arbitrary solutions for all $\,\a_{(1,i,\ell_i)}$. In this case, all the
  vectors in the chain are tangent.

  \noindent $\bullet$ Consider finally a chain\, $\{v_{(\la,i,\ell_i)}\}$
  \,relative to an eigenvalue $\la\neq \pm
  1$. Eqs.~(\ref{tan-1})-(\ref{tan-2}) give arbitrary solutions for all
  $\,a_{(\la,i,\ell_i)}$ and again all the vectors in the chain occur in
  $\pa_\tau\/x$.
\par}
\noindent The space $\Pi_g$ of tangent directions is then
\begin{equation*}
     \Pi_g=\textnormal{Span}\,\big\{v_{(\la,i,\ell_i)}\!:~ 
              (\la,\ell_i)\neq (-1, L_i)\,\big\}
\end{equation*}
and has dimension ${\tt d}-m_{-1}$, where we recall that ${\tt d}$ is the
group dimension and $m_{-1}$ the geometric multiplicity of $\la=-1$. Since the
nontangent vectors $v_{(-1,i,L_i)}$ are removed from the set of all
(generalized) eigenvectors through the action of ${\cal F}+1$, one has
\begin{equation}
  \Pi_g = ({\cal F}+1)\,\textnormal{Span}\,\big\{v_{(\la,i,\ell_i)}\big\}  \,
	=  \textnormal{Im}\big( {\cal F} + 1 \big)\,.
\label{t-directions}
\end{equation}
This is precisely eq.~(\ref{Pi-g}).

\section*{Appendix B. Metrically degenerate tangent planes}
\renewcommand{\theequation}{B.\arabic{equation}}
\setcounter{equation}{0}

Here we explicitely construct tangent vectors that are orthogonal to all
tangent vectors, including itself, so they define a metrically degenerate
tangent plane $\Pi_g$.

To this end, we first note that the isometry property~(\ref{isometry}) and
eq.~(\ref{chain}) imply the orthogonality relation
\begin{equation}
  (1-\la\mu)\;\mG\big(\,{v}_{(\la\,,\,i\,,\,\ell_i)}\,, 
        \,{v}_{(\mu\,,\,j\,,\,m_j)}\,\big)=0   \,.
\label{orthogonality}
\end{equation}
for two arbitrary (generalized) eigenvectors.

Assume for concreteness that there is only one chain\, $\{v_{-1,1,1}\ldots\/
v_{(-1,1,L)}\}$ \,of\, $L\!\geq\! 2$ \,generalized eigenvectors relative to
the eigenvalue $\la=\!-\,1$, and let us write $u_\ell:=v_{(-1,1,\ell)}$ for its
members. As explained in Appendix A, the first $L-1$ vectors in this chain
define directions in~$\,\Pi_g$. Noting that ${\cal F}$ is an isometry and
recalling eqs.~(\ref{chain}), we have
\begin{equation*}
     \mG \big( u_1\,,\,u_{\ell+1}\big)    
         = \mG \big({\cal F}u_1\,,\,{\cal F}u_{\ell+1}\big) 
          = \mG \big( u_1\,,\,u_{\ell+1}\,\big)
               -\mG \big( u_1\,,\,u_{\ell}\,\big)
  \qquad \ell=1,\ldots,L-1\,.
\end{equation*}
It follows that\, $\mG \big( u_1,u_{\ell}\big)=0$ \,for\, $\ell=1\ldots,L-1$.
Since $\{u_\ell\}$ is the only chain with eigenvalue $-1$, any other direction
in $\Pi_g$ has eigenvalue $\la\neq\!-1$, and thus eq.~(\ref{orthogonality})
implies that it is orthogonal to\, $u_1$. The eigenvector\, $u_1$ is thus
orthogonal to all (generalized) eigenvectors spanning $\Pi_g$, and in
particular to itself.

It is trivial to extend these arguments to show that every eigenvector heading
a chain with eigenvalue $\la=\!-\,1$ defines a null direction orthogonal to
$\Pi_g$.

\section*{Appendix C. Invariance of the spectrum of
  $\textnormal{Ad}_{g^{-1}}R$}
\renewcommand{\theequation}{C.\arabic{equation}}
\setcounter{equation}{0}

This Appendix contains the discussion of the invariance of the spectrum of the
operator $\textnormal{Ad}_{g^{-1}}R$ under \hbox{$R$-twined} conjugation,
where $R$ is a Lie algebra automorphism.

The eigenvalue problem for $\textnormal{Ad}_{g^{-1}}R$ takes the form
\begin{equation}
     R\,V_{(\la,\ell)}g=\la g\,V_{(\la,\ell)}+g\,V_{(\la,\ell-1)} \,,
\label{specR}
\end{equation}
where the last term accounts for the occurrence of generalized eigenvectors.
Here the chain labeling index $i$ in Appendices A and B has been omitted in
the notation since it does not play any r\^ole. An arbitrary \hbox{$R$-twined}
conjugate~$g^{\,\prime}$ of~$g$ can be written as
\begin{equation*}
		g^{\,\prime}=e^{\,- RU}g\,e^{U} \,,
\end{equation*}
for some $U$ in $\ag$. After some trivial manipulations, eq.~(\ref{specR}) can
be written in terms of $g'$ as
\begin{equation}
  e^{-RU}\,RV_{(\la,\ell)}\,e^{\,RU}\,g^{\,\prime}
    = \la\, g^{\,\prime} e^{-U}\,V_{(\la,\ell)}\,e^{\,U}
    + g^{\,\prime} e^{-U}\,V_{(\la,\ell-1)}\,e^{\,U}\,.
\label{specR2}
\end{equation}
Being $R$ a Lie algebra automorphism, the left hand side of this equation is\,
$R\big(e^{-U} V_{(\la,\ell)} e^{\,U}\big)g^{\,\prime}$. Eq.~(\ref{specR2})
becomes then
\begin{equation*}
    \textnormal{Ad}_{g'^{-1}}R\,V'_{(\la,\ell)} 
        = \la V'_{(\la,\ell)}+V'_{(\la,\ell-1)} ~~\qquad
    V^{\,\prime}_{(\la,\ell)}= e^{-U} V_{(\la,\ell-1)} e^{\,U} \,.
\end{equation*}
The eigenvalues thus remain invariant, while the (generalized) eigenvectors
change by ordinary conjugation. As a consequence, the dimension of the
linear space generated by the eigenvectors associated to a given eigenvalue is
constant under \hbox{$R$-twined} conjugation.
        
\section*{Acknowledgment}

The authors are grateful to C.~Moreno for conversations, and to MEC and
UCM-BSCH, Spain for partial support through grants FPA2008-04906 and
910770-GR35/10-A.

\end{document}